\documentclass[aps,prl,reprint,superscriptaddress]{revtex4-2}
\usepackage{amssymb,amsmath,bm,braket,txfonts,graphicx,wasysym,tikz}
\usepackage[bookmarks=true,colorlinks,citecolor=blue,urlcolor=blue]{hyperref}
\newcommand{\plaqa}{\tikz[baseline=-0.62ex]{\draw [black] (0.0,0.0) -- (0.15,0.25); \draw [black] (0.15,0.25) -- (0.45,0.25); \draw [black] (0.45,0.25) -- (0.6,0.0); \draw [black] (0.6,0.0) -- (0.45,-0.25); \draw [black] (0.45,-0.25) -- (0.15,-0.25); \draw [black] (0.15,-0.25) -- (0.0,0.0); \filldraw [color=black, fill=black!100] (0,0) circle (0.05 and 0.05); \filldraw [color=black, fill=black!25] (0.45,0.25) circle (0.05 and 0.05); \filldraw [color=black, fill=black!0] (0.45,-0.25) circle (0.05 and 0.05);}}
\newcommand{\plaqb}{\tikz[baseline=-0.62ex]{\draw [black] (0.0,0.0) -- (0.15,0.25); \draw [black] (0.15,0.25) -- (0.45,0.25); \draw [black] (0.45,0.25) -- (0.6,0.0); \draw [black] (0.6,0.0) -- (0.45,-0.25); \draw [black] (0.45,-0.25) -- (0.15,-0.25); \draw [black] (0.15,-0.25) -- (0.0,0.0); \filldraw [color=black, fill=black!100] (0.15,0.25) circle (0.05 and 0.05); \filldraw [color=black, fill=black!25] (0.6,0.0) circle (0.05 and 0.05); \filldraw [color=black, fill=black!0] (0.15,-0.25) circle (0.05 and 0.05);}}
\newcommand{\plaqc}{\tikz[baseline=-0.62ex]{\draw [black] (0.0,0.0) -- (0.15,0.25); \draw [black] (0.15,0.25) -- (0.45,0.25); \draw [black] (0.45,0.25) -- (0.6,0.0); \draw [black] (0.6,0.0) -- (0.45,-0.25); \draw [black] (0.45,-0.25) -- (0.15,-0.25); \draw [black] (0.15,-0.25) -- (0.0,0.0); \filldraw [color=black, fill=black!100] (0.15,0.25) circle (0.05 and 0.05); \filldraw [color=black, fill=black!25] (0.6,0.0) circle (0.05 and 0.05); \filldraw [color=black, fill=black!0] (0.15,-0.25) circle (0.05 and 0.05);}}
\newcommand{\plaqd}{\tikz[baseline=-0.62ex]{\draw [black] (0.0,0.0) -- (0.15,0.25); \draw [black] (0.15,0.25) -- (0.45,0.25); \draw [black] (0.45,0.25) -- (0.6,0.0); \draw [black] (0.6,0.0) -- (0.45,-0.25); \draw [black] (0.45,-0.25) -- (0.15,-0.25); \draw [black] (0.15,-0.25) -- (0.0,0.0); \filldraw [color=black, fill=black!0] (0,0) circle (0.05 and 0.05); \filldraw [color=black, fill=black!100] (0.45,0.25) circle (0.05 and 0.05); \filldraw [color=black, fill=black!25] (0.45,-0.25) circle (0.05 and 0.05);}}

\begin{document}

\def\papertitle{{Unconventional spin transport in strongly correlated kagome systems}}
\def\tum{{Technical University of Munich, TUM School of Natural Sciences, Physics Department, 85748 Garching, Germany}}
\def\mcqst{{Munich Center for Quantum Science and Technology (MCQST), Schellingstr. 4, 80799 M{\"u}nchen, Germany}}
\newcommand{\TUM}{\affiliation{\tum}}
\newcommand{\MCQST}{\affiliation{\mcqst}}
\title{\papertitle}
\author{Masataka Kawano} \TUM
\author{Frank Pollmann} \TUM \MCQST
\author{Michael Knap} \TUM \MCQST
\date{\today}

\begin{abstract}
Recent progress in material design enables the study of correlated, low-temperature phases and associated anomalous transport in two-dimensional kagome systems. Here, we show that unconventional spin transport can arise in such systems even at elevated temperatures due to emergent dynamical constraints. To demonstrate this effect, we consider a strong-coupling limit of an extended Hubbard model on the kagome lattice with density of $2/3$. We numerically investigate the charge and spin transport by a cellular automaton circuit, allowing us to perform simulations on large systems to long times while preserving the essential conservation laws. The charge dynamics reflects the constraints and can be understood by a Gaussian field theory of a scalar height field. Moreover, the system exhibits a hidden spin conservation law with a dynamic sublattice structure, which enables additional slow relaxation pathways for spin excitations. These features can be directly tested by measuring the dynamic spin structure factor with neutron scattering.
\end{abstract}
\maketitle

\emph{\textbf{Introduction.---}}Over the recent years different material classes of kagome systems have been identified. In these systems, a quasi two-dimensional lattice of corner-shared triangles is formed with bands that are partially filled. The unique geometric structure of the kagome lattice gives rise to a rich variety of charge- and spin-ordered low-temperature states, leading to unconventional transport phenomena. Previous studies have predominantly focused on transport in these exotic phases of matter. Examples include the emergence of giant anomalous Hall and Nernst effects in noncollinear antiferromagnetic kagome materials Mn$_{3}X$ ($X=$Sn, Ge, Ga)~\cite{nakatsuji2015nature,nayak2016sciad,kiyohara2016prap,liu2017srep,ikhlas2017nphys, kuroda2017nmat,yang2017njp}, collinear kagome antiferromagnets Fe$X$ ($X=$Sn,Ge)~\cite{kang2020nmat,xie2021comphys,sales2019prm,teng2022nature}, and kagome ferromagnets such as Fe$_3$Sn$_2$~\cite{wang2016prb,ye2018nature,hou2017advmat,hou2018nlet} and Co$_3$Sn$_2$S$_2$~\cite{wang2018ncom,liu2019science,guguchia2020ncom}. Recently, a giant anomalous Hall effect has been observed in a charge density wave phase of the kagome metal $A$V$_{3}$Sb$_{5}$ ($A=$K, Rb, Cs) as well~\cite{yang2020sciad,yu2021prb}. At even lower temperatures, these materials exhibit superconductivity~\cite{ortiz2019prm,ortiz2020prl,ortiz2021prm,yin2021crl}, further highlighting their unique electronic properties. Moreover, at certain commensurate fillings, an exotic interplay of charge and spin fluctuations has been theoretically predicted to arise in kagome systems~\cite{Pollmann2008, pollmann2014prb, Ferhat2014}. There, emergent constraints arise from the combined effect of strong interactions and geometric frustration, which leads to exotic ground states at low temperatures~\cite{Pollmann2008, pollmann2014prb, Ferhat2014}.

\begin{figure}[t]
    \includegraphics[width=85mm]{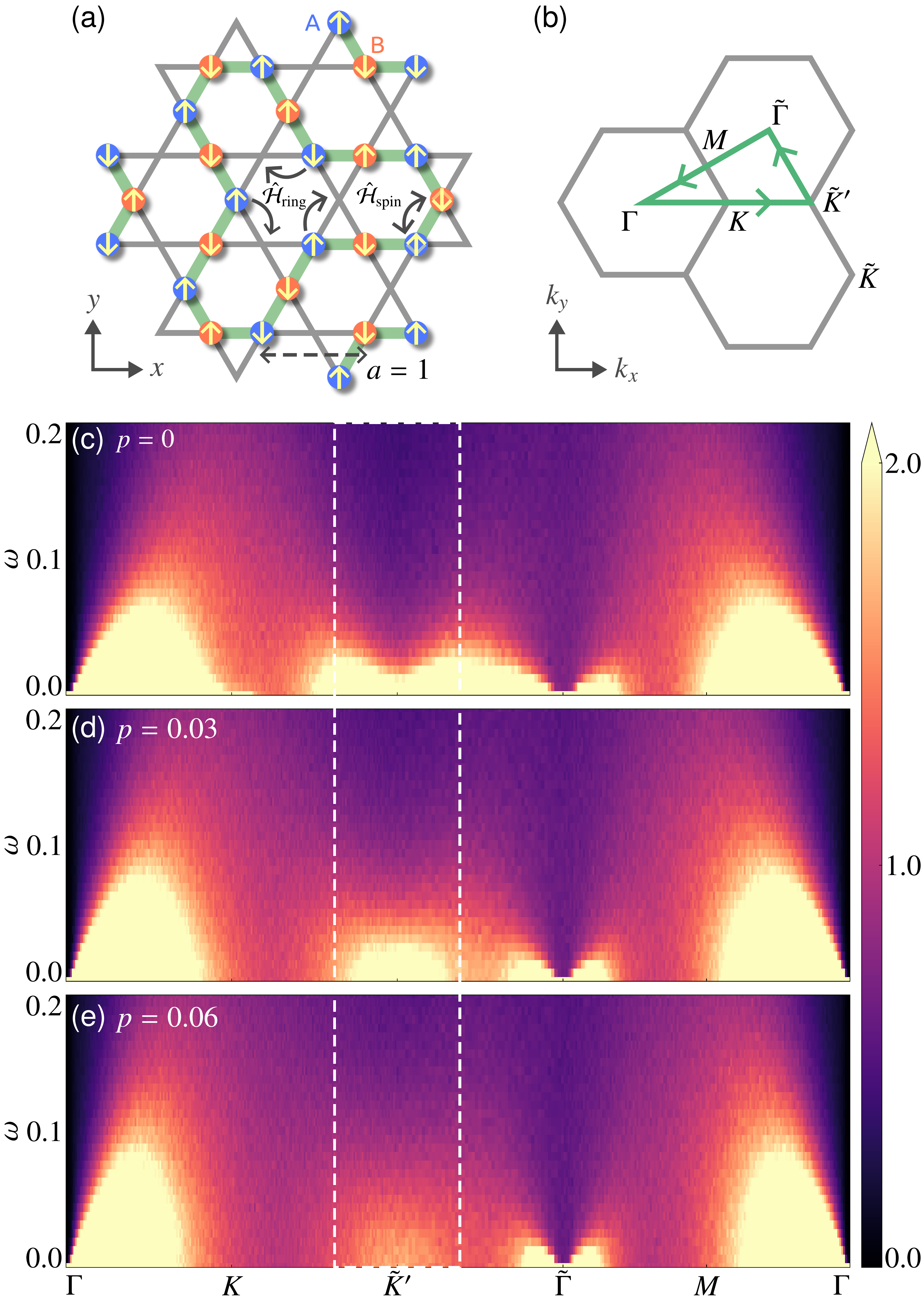}
    \caption{\textbf{Constrained dynamics on the kagome lattice.} We study the high-temperature transport of a kagome system with density $2/3$ in the strong coupling. (a) Snapshot of a charge and spin configuration. The dynamics is governed by ring-exchange processes $\hat{\mathcal{H}}_{\mathrm{ring}}$ and a small fraction $p$ of spin-exchange processes $\hat{\mathcal{H}}_{\mathrm{spin}}$. The dynamics exhibits a hidden sublattice structure, illustrated by blue and orange circles on dynamical sublattices A and B. In the absence of spin exchange $p=0$, the total spin on each dynamic sublattice is conserved. (b) Extended Brillouin zone of the kagome lattice. (c-e) Dynamic spin structure factor $S(\bm{k},\omega)$ along the cuts shown in (b) with (c) $p=0$, (d) $p=0.03$, and (e) $p=0.06$. The hidden spin conservation law manifests itself as a strong spectral response near the $\tilde K'$ point (enclosed by the white dashed line) and decreases with increasing spin exchange $p$.}
    \label{fig:sdsf}
\end{figure}

Here, we show that the emergent constraints lead to unconventional charge and spin transport even in the disordered, high-temperature phase of a strongly correlated kagome system. We numerically study transport with cellular automaton circuits that encode the constraints exactly and allow us to simulate large system sizes and long time scales~\cite{medenjak2017prl,sarang2018qst,iaconis2019prb,feldmeier2020prl}. The charge dynamics directly reflects the constraints and can be explained with an effective field theory for a height variable. Moreover, we find that the spin dynamics exhibits additional slow relaxation pathways due to an emergent constraint with a dynamic sublattice structure. This emergent constraint acts as a spatially modulated symmetry of spins, leading to an additional slow mode at finite momenta. It arises from the interplay between strong electron correlations and geometric frustration of the kagome lattice. We demonstrate that the unconventional spin transport appears as a characteristic spectral feature in a dynamic spin structure factor near the $\tilde K'$ point, which can be experimentally measured by inelastic neutron scattering as shown in Fig.~\ref{fig:sdsf}.

\emph{\textbf{Model.---}}We consider an extended Hubbard model on the kagome lattice with density of $2/3$,
\begin{align}
    \hat{\mathcal{H}} &= -\mathfrak{t}\sum_{\braket{\bm{r},\bm{r}'}}\sum_{\sigma}\left(\hat{c}_{\bm{r},\sigma}^{\dagger}\hat{c}_{\bm{r}',\sigma}+\mathrm{H.c.}\right) \nonumber \\
                      &\hspace{20pt} + U\sum_{\bm{r}}\hat{n}_{\bm{r},\uparrow}\hat{n}_{\bm{r},\downarrow} + V\sum_{\braket{\bm{r},\bm{r}'}}\hat{n}_{\bm{r}}\hat{n}_{\bm{r}'},
    \label{eq:H}
\end{align}
where $\braket{\bm{r},\bm{r}'}$ denotes nearest-neighbors, $\mathfrak{t}$ is the hopping amplitude, $U$ and $V$ ($U>V$) are the on-site and nearest-neighbor repulsive interactions, $\hat{c}_{\bm{r},\sigma}$ ($\hat{c}_{\bm{r},\sigma}^{\dagger}$) annihilates (creates) an electron at site $\bm{r}$ with spin $\sigma=\uparrow,\downarrow$, $\hat{n}_{\bm{r}}=\sum_{\sigma}\hat{n}_{\bm{r},\sigma}$ is the particle density operator at site $\bm{r}$, and $\hat{n}_{\bm{r},\sigma}=\hat{c}_{\bm{r},\sigma}^{\dagger}\hat{c}_{\bm{r},\sigma}$. In the strong-coupling limit, $U,V\gg\mathfrak{t}$, large on-site interactions $U$ project out double occupancies. Then the low-energy manifold has exactly two electrons per triangle due to large nearest-neighbor interactions $V$ as shown in Fig.~\ref{fig:sdsf}(a). Treating the nearest-neighbor hopping $\mathfrak{t}$ as a perturbation, we obtain a lowest order effective Hamiltonian $\hat{\mathcal{H}}_{\mathrm{eff}} = \hat{\mathcal{H}}_{\mathrm{ring}} + \hat{\mathcal{H}}_{\mathrm{spin}}$~\cite{pollmann2014prb}. The first term describes ring exchange processes
\begin{align}
    \hat{\mathcal{H}}_{\mathrm{ring}} = -g\sum_{\hexagon}\left(\Ket{\plaqa}\Bra{\plaqb}+\Ket{\plaqc}\Bra{\plaqd}+\mathrm{H.c.}\right),
    \label{eq:Hring}
\end{align}
where $g=6\mathfrak{t}^{3}/V^{2}$ is the coupling constant and $\sum_{\hexagon}$ runs over all hexagons on the kagome lattice. $\hat{\mathcal{H}}_{\mathrm{ring}}$ consists of the collective clockwise and counterclockwise motions of three electrons on the hexagons, as illustrated in Fig.~\ref{fig:sdsf}(a). The second term $\hat{\mathcal{H}}_{\mathrm{spin}}$ is the nearest-neighbor Heisenberg exchange interaction
\begin{align}
    \hat{\mathcal{H}}_{\mathrm{spin}} = J\sum_{\braket{\bm{r},\bm{r}'}}\left(\hat{\bm{S}}_{\bm{r}}\cdot\hat{\bm{S}}_{\bm{r}'}-\frac{1}{4}\hat{n}_{\bm{r}}\hat{n}_{\bm{r}'}\right),
    \label{eq:Hspin}
\end{align}
where $\hat{S}_{\bm{r}}^{\mu}=(1/2)\sum_{\sigma,\sigma'}\hat{c}_{\bm{r},\sigma}^{\dagger}\tau_{\sigma,\sigma'}^{\mu}\hat{c}_{\bm{r},\sigma'}$ is the spin operator at site $\bm{r}$ with Pauli matrix $\tau^{\mu}$ ($\mu=x,y,z$) and $J=4\mathfrak{t}^{2}/(U-V)+(4\mathfrak{t}^{3}/V^2)\cdot(U+V)/(U-V)$ is the exchange coupling constant.

The system we consider conserves both the total charge and total spin. Therefore, we will investigate in this work both charge and spin transport. In addition to these conservation laws, for $J=0$ the system has two unconventional conservation laws due to the characteristic motion of the electrons introduced by $\hat{\mathcal{H}}_{\mathrm{ring}}$~\cite{pollmann2014prb}: First, a total charge conservation law on any straight line along the edge of a hexagon. Second, an emergent conservation law of spins on dynamically moving sublattices. To visualize this conservation law, draw loops by connecting nearest-neighbor electrons. Every loop consists of an even-number of sites. Therefore, we can assign sublattice labels, A and B, to each electron in the loop; see orange and blue circles in Fig.~\ref{fig:sdsf} (a). We choose the sublattice labels such that the next-nearest-neighbor electrons have the same label. Then each electron loop has a bipartite structure which is conserved during the time evolution. It has, however, dynamics associated with it as the electrons move by the ring exchange processes. Hence, we call it a \textit{dynamic sublattice}. Since spins on different dynamic sublattices never exchange with each other for $J=0$, which can be achieved by tuning $\mathfrak{t}$, $U$, and $V$ appropriately~\cite{pollmann2014prb}, the total spin on each dynamic sublattice is another conserved quantity in this limit.

We are interested in studying the consequences of the interplay between the conservation laws at high temperatures $U,V\gg k_{\mathrm{B}}T\gg\mathfrak{t}$. In this regime, the high-energy sector violating the constraint of two electrons per triangle does not contribute, and the dynamics is determined by the low-energy effective Hamiltonian $\hat{\mathcal{H}}_{\mathrm{eff}}$. The late-time transport is governed by classical hydrodynamics which respects the structure of the conservation laws~\cite{chaikin1995book,mukerjee2006prb,lux2014pra,bohrdt2017njp}. System-specific details only enter in the value of the transport coefficients, i.e., the diffusion constants, which we do not aim to determine quantitatively. This allows us to study the dynamics of our system by classically simulable cellular automaton circuits, which respect the conservation laws of the system~\cite{medenjak2017prl,sarang2018qst,iaconis2019prb,feldmeier2020prl}. The cellular automaton circuit describes the discrete time evolution which is designed so that a product state is mapped to another product state at each time step. A single time step in our cellular automaton circuit consists either of $\hat{U}_{\mathrm{ring}}$ or $\hat{U}_{\mathrm{spin}}$, which capture local ring-exchange processes and spin flips, mimicking the action of $\hat{\mathcal{H}}_{\mathrm{ring}}$ and $\hat{\mathcal{H}}_{\mathrm{spin}}$, respectively; see Supplemental Materials for details~\cite{supp}. The unitary operator for the automaton time evolution is then given by a sequential action, $\hat{U}(t)=\cdots\hat{U}_{\mathrm{ring}}\hat{U}_{\mathrm{ring}}\hat{U}_{\mathrm{spin}}\hat{U}_{\mathrm{ring}}\cdots$. The effect of finite $J/g$ is reflected as a finite probability $p$ of applying $\hat{U}_{\mathrm{spin}}$ at each time step, which we use as a tuning parameter for the dynamics. The cellular automaton circuit with small $p$ effectively describes the system's dynamics with small $|J/g|$~\cite{supp}. The time evolution of correlation functions are calculated from the sequence of the product states generated by the cellular automaton time evolution. Numerical simulations are performed for systems with $150\times150$ hexagons with periodic boundary conditions. We have checked that on the considered time scales finite size effects are absent~\cite{supp}. The correlation functions are averaged over $2\times10^{5}$ random initial product states.

\begin{figure}[t]
    \includegraphics[width=85mm]{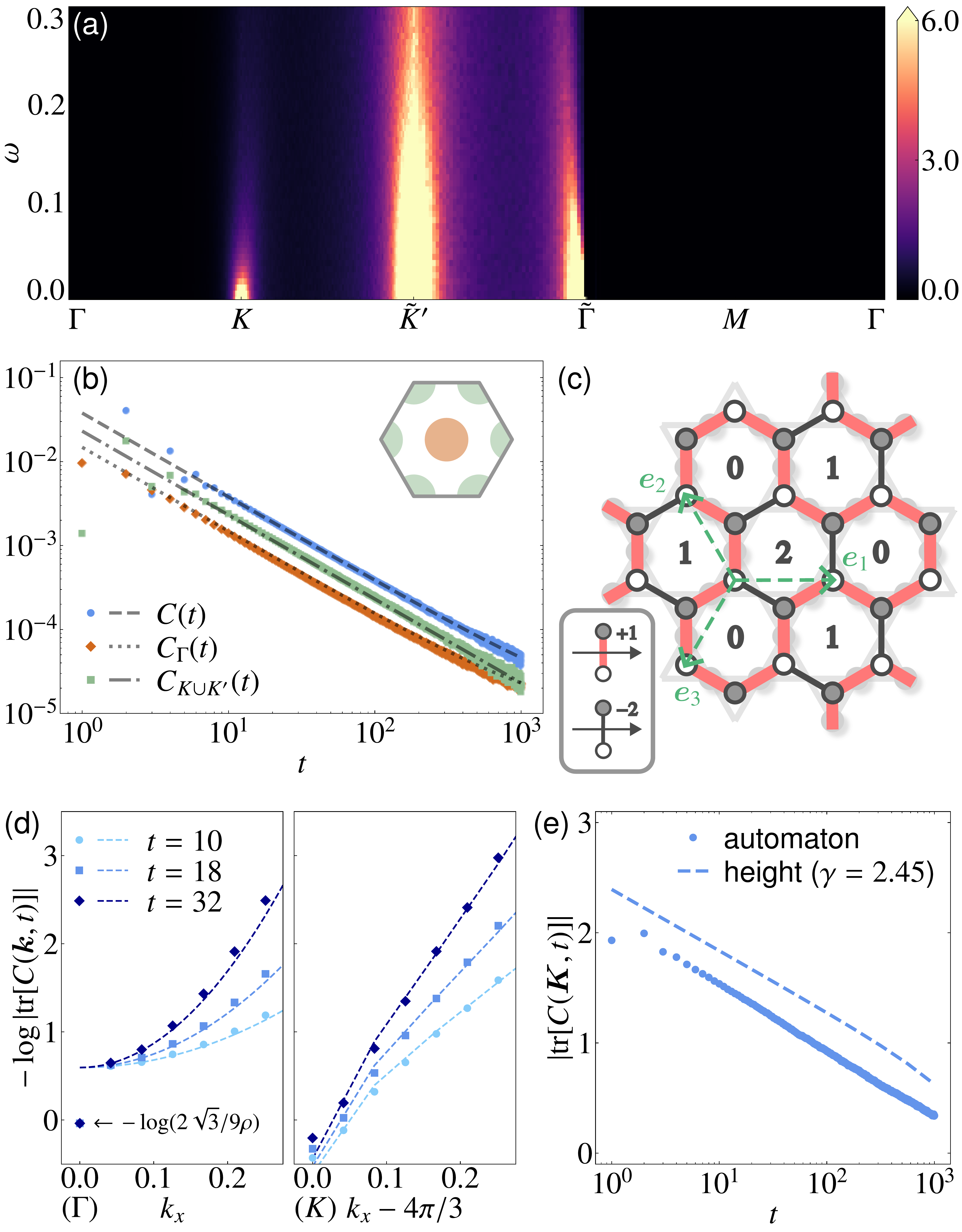}
    \caption{\textbf{Charge relaxation dynamics for $\bm{p=0}$.} (a) Dynamic charge structure factor $C(\bm{k},\omega)$. (b) The charge autocorrelation function obtained numerically (dots) is compared to the height-field theory (dashed lines), which is almost identical to $t^{-1}$. Inset: Summation range of $C_{\Gamma}(t)$ and $C_{K\cup K'}(t)$. (c) The charge configuration maps onto a fully packed loop model on the dual honeycomb lattice which enables one to introduce a height variable whose value is indicated in each hexagon. (d) Time-dependent charge structure factor at $t=10,18,32$ near the $\Gamma$ (left) and $K$ point (right) along the $k_{x}$ direction (blue dots). Good agreement is found with the height-field theory (dashed line). (e) Time evolution of $|\mathrm{tr}[C(\bm{K},t)]|$ obtained from the cellular automaton (dots) and height-field theory (dashed line) at the $K$ point. We set $\gamma=2.45$ throughout, which is the free parameter of the height field theory.}
    \label{fig:charge}
\end{figure}

\emph{\textbf{Charge relaxation dynamics.---}}We first study the charge dynamics by measuring the dynamic charge structure factor
\begin{align}
    C(\bm{k},\omega) = \frac{1}{N}\sum_{\bm{r},\bm{r}'}\int_{-\infty}^{\infty}dt\ C_{\bm{r},\bm{r}'}(t)\mathrm{e}^{-i\bm{k}\cdot(\bm{r}-\bm{r}')+i\omega t},
    \label{eq:cdsf}
\end{align}
where $N$ is the number of sites and $C_{\bm{r},\bm{r}'}(t)$ is the charge correlation function defined as
\begin{align}
    C_{\bm{r},\bm{r}'}(t) = \braket{\delta\hat{n}_{\bm{r}}(t)\delta\hat{n}_{\bm{r}'}(0)}.
\end{align}
Here, $\delta\hat{n}_{\bm{r}}(t)=\sum_{\sigma}\hat{n}_{\bm{r},\sigma}(t)-2/3$ specifies the deviation from the average charge density at time $t$ and $\braket{\cdots}$ is taken over an infinite-temperature ensemble of charge and spin configurations fulfilling the constraints. We use the worm algorithm to sample these configurations~\cite{rahman1972jcp,evertz1993prl}. Cellular automaton results for the dynamic charge structure factor obtained with $p=0$ ($J=0$) are shown in Fig.~\ref{fig:charge}(a). Since the kagome lattice has a three-sublattice structure, the structure factor has to be considered on the extended Brillouin zone shown in Fig.~\ref{fig:sdsf}(b). The dynamic charge structure factor carries spectral weight only near the $K$, $\tilde{K}'$, and $\tilde{\Gamma}$ points but no weight is observed near the $\Gamma$ point. The asymmetry of the spectral response near $\tilde{\Gamma}$ can be understood from an underlying pinch point~\cite{moessner2003prb}. The absence of the spectral weight near the $\Gamma$ point results from the local constraint of having exactly two electrons per triangle $\sum_{\bm{r}\in\triangle}\delta\hat{n}_{\bm{r}}(t)=0$~\cite{moessner2003prb}.

To gain further insight into the charge dynamics, we examine the charge autocorrelation function $C(t)=(1/N)\sum_{\bm{r}}C_{\bm{r},\bm{r}}(t)=(1/N)\sum_{\bm{k}}\mathrm{tr}[C(\bm{k},t)]$ and its momentum-resolved contributions, $C_{\Gamma}(t)=(1/N)\sum_{\bm{k}\sim\bm{0}}\mathrm{tr}[C(\bm{k},t)]$ and $C_{K\cup K'}(t)=(1/N)\sum_{\bm{k}\sim\bm{K},\bm{K}'}\mathrm{tr}[C(\bm{k},t)]$, where $C(\bm{k},t)$ is the time-dependent Fourier component defined as
\begin{align}
    C_{\ell,\ell'}(\bm{k},t) &= \frac{1}{N/3}\sum_{\bm{r}\in I_{\ell}}\sum_{\bm{r}'\in I_{\ell'}}C_{\bm{r},\bm{r}'}(t)\mathrm{e}^{-i\bm{k}\cdot(\bm{r}-\bm{r}')},
    \label{eq:tdc-fourier}
\end{align}
where $I_{\ell}$ is the set of sites on the sublattice $\ell=1,2,3$ of the kagome lattice.

In the hydrodynamic limit, the relaxation dynamics of systems with a simple charge conservation law is described by the diffusion equation $\partial_t n = D \nabla^{2}n$, where $n$ is a coarse-grained charge density and $D$ is a diffusion constant. From that the correlation function can be obtained, which in two-dimensions is given by $C^\text{D}_{\bm r, \bm r'}(t) = (1/4\pi Dt)\exp\left[-|\bm r-\bm r'|^2/4Dt\right]$. The numerically computed autocorrelation function of our kagome system, evaluated at $\bm r = \bm r'$, decays as $C(t)\sim t^{-1}$, see Fig.~\ref{fig:charge}(b), which is consistent with the prediction from conventional hydrodynamic diffusion. However, due to the underlying constraints, the autocorrelation function has strong contributions from the $\bm{K}=(4\pi/3,0)$ and $\bm{K}'=(2\pi/3,2\pi/\sqrt{3})$ points, which are the wave vectors at the corner of the Brillouin zone. This is in stark contrast to the result obtained from the simple diffusion equation, which upon Fourier transform leads to an exponential decay for finite-momentum correlations $C^\text{D}(\bm k,t) = \exp(- D {\bm k}^2 t)$.

To understand how this unconventional charge relaxation emerges from the constraints, we utilize an effective Gaussian field theory using a height representation of the dual loop model~\cite{henley1997jsp,zeng1997prb,moessner2003prb}. Any charge configuration of the system can be mapped to the fully packed loop configuration on the dual honeycomb lattice as shown in Fig.~\ref{fig:charge}(c), in which loop segments are occupied when the particles sit on the links of the dual lattice~\cite{pollmann2014prb}. The integer-valued height variable is assigned to each hexagon through the following rule: the height increases by one if a loop segment is crossed and decreases by two otherwise when moving counterclockwise around a gray honeycomb site in Fig.~\ref{fig:charge}(c). When taking the continuum limit of the height variable, the constraints have to be carefully considered, which leads to low energy modes near the $\Gamma$, as well as the $K$ and $K'$ points. The dynamical charge correlations can then be obtained analytically by assuming that the dynamics is governed by a Langevin equation; we present the details of this approach following Refs.~\cite{henley1997jsp, moessner2003prb} in Supplemental material~\cite{supp}. We compare the predictions for the autocorrelations obtained from the height field theory in Fig.~\ref{fig:charge}(b) and find excellent agreement with the cellular automaton simulation. When comparing the field-theoretic predictions with our numerical simulations, we only need to specify the effective diffusion constant, which is the same for all comparisons shown in Fig.~\ref{fig:charge}~\cite{supp}.

To further substantiate the effective height field representation of the charge dynamics, we compare the  time-dependent Fourier component of the charge correlations $\mathrm{tr}[C(\bm{k},t)]$ near $\Gamma$ and $K$ points along $k_{x}$ direction. The field theory and numerical results exhibit excellent agreement except precisely at the $K$ point; Fig.~\ref{fig:charge}(e). This deviation may arise from subleading terms that are relevant only at the $K$ and $K'$ points, which would be interesting to identify in future work. Nevertheless, the main characteristics of the charge dynamics can be understood from the height field theory. The additional contributions to the charge correlations around the $K$ and $K'$ points are thus the consequence of the local charge constraints that arise in the strong coupling limit.

\emph{\textbf{Spin relaxation dynamics.---}}We now turn to the dynamic spin structure factor
\begin{align}
    S(\bm{k},\omega) &= \frac{1}{N}\sum_{\bm{r},\bm{r}'}\int_{-\infty}^{\infty}dt\ S_{\bm{r},\bm{r}'}(t)\mathrm{e}^{-i\bm{k}\cdot(\bm{r}-\bm{r}')+i\omega t},
    \label{eq:sdsf}
\end{align}
where $S_{\bm{r},\bm{r}'}(t) = \braket{\hat{S}_{\bm{r}}^{z}(t)\hat{S}_{\bm{r}'}^{z}(0)}$. From the conventional diffusion equation, the correlation function in frequency and momentum space is $C^\text{D}(\bm{k},\omega) \sim (D\bm{k}^{2})/((D\bm{k}^{2})^{2}+\omega^{2})$. Thereby fixed-momentum cuts exhibit a peak at $\omega=0$ with height $(1/D\bm{k}^{2})$ and width $D\bm{k}^{2}$, displaying a dome-like spectral distribution near $\Gamma$ point~\cite{vineyard1958pr,hollander1994jsp}. Figure~\ref{fig:sdsf}(c) shows the dynamic spin structure factor with $p=0$ ($J=0$). We observe a broad spectral distribution near $\tilde{K}'$ point in addition to the conventional dome-like structures near $\Gamma$ and $\tilde{\Gamma}$ points. However, for conventional spin-diffusion, no response is expected near $\tilde{K}'$ point.

To pinpoint the origin of the additional spectral intensity near $\tilde{K}'$ point, we perform additional simulations with finite spin exchange probability $p$ ($J\neq0$), which breaks the spin conservation law on the dynamic sublattices. As a consequence for sufficiently large $p$, the dynamics is expected to follow $C^\text{D}(\bm{k},\omega)$, and the response near the $\tilde K'$ point to vanish. Our simulations indeed confirm this expectation; see Fig.~\ref{fig:sdsf}(d) and~\ref{fig:sdsf}(e) for $p=0.03$ and $p=0.06$. The larger the spin-exchange probability, the weaker are the features near the $\tilde{K}'$ point. From that we can conclude, that the additional contribution observed in the dynamic spin structure factor in Fig.~\ref{fig:sdsf}(c) results from the spin conservation law on the dynamic sublattices. We also show that adding next-nearest-neighbor spin exchange terms, which respects the spin conservation law on the dynamic sublattices, preserves the sharp spectral features near the $\tilde{K}'$ point; see Supplemental material~\cite{supp}.

\begin{figure}[t]
    \includegraphics[width=85mm]{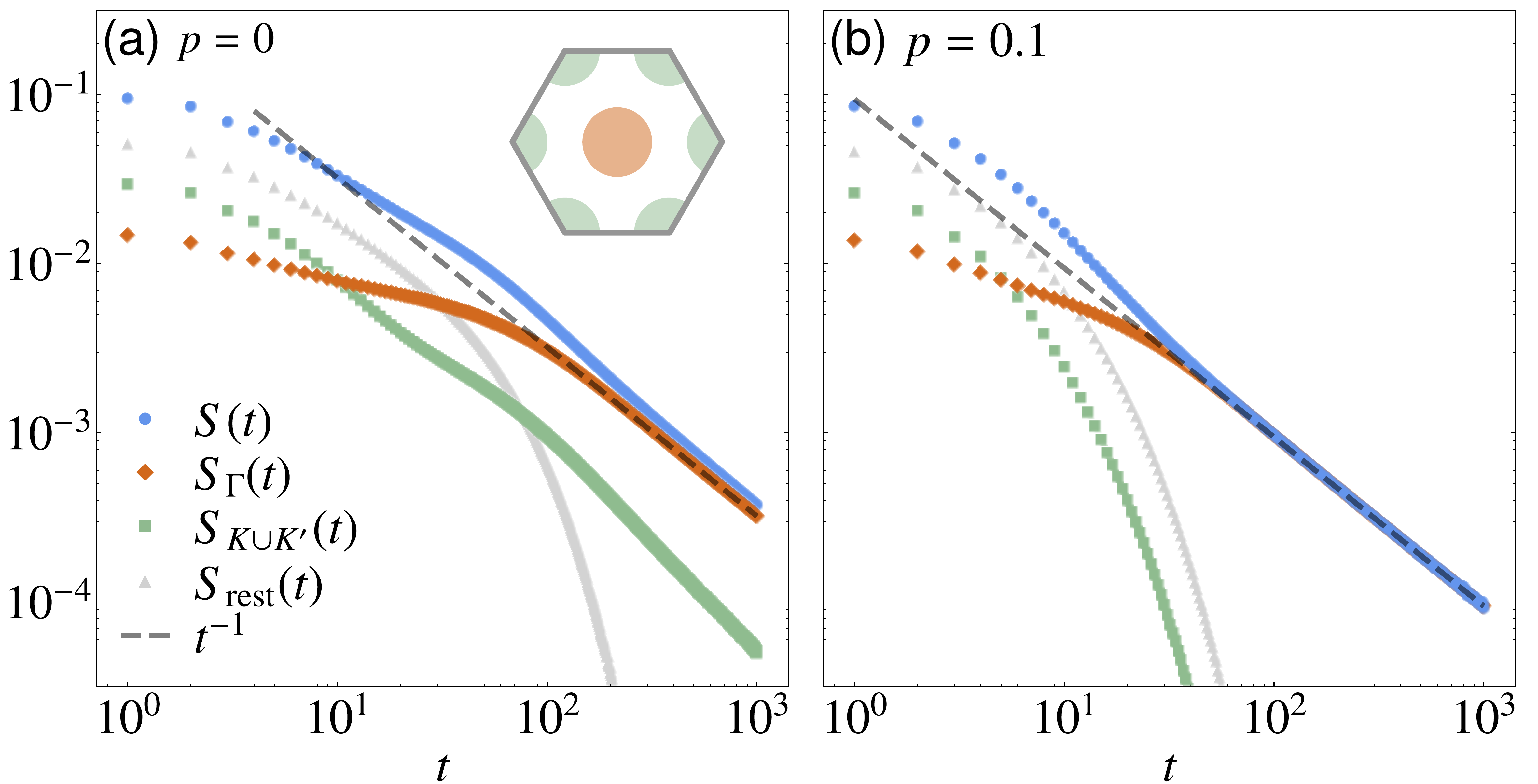}
    \caption{\textbf{Spin autocorrelation function.} Time evolution of the spin autocorrelation function $S(t)$ with (a) $p=0$ and (b) $p=0.1$. The momentum-resolved contributions, $S_{\Gamma}(t)$, $S_{K\cup K'}(t)$, and $S_{\mathrm{rest}}(t)=S(t)-S_{\Gamma}(t)-S_{K\cup K'}(t)$, are also shown. The inset in (a) shows their summation range in the first Brillouin zone. Although $S(t)$ finally reaches the diffusive scaling $t^{-1}$ (gray dashed line), there is still large anomalous regime in (a) due to the finite $S_{K\cup K'}(t)$, arising from the spin conservation law on the dynamical sublattices.}
    \label{fig:st}
\end{figure}

In order to gain further insights into the spin dynamics, we investigate the spin autocorrelation function $S(t)=(1/N)\sum_{\bm{r}}S_{\bm{r},\bm{r}}(t)$. For systems with a trivial spin conservation law, the autocorrelation is again expected to follow the diffusive scaling $S(t)\sim t^{-1}$ in two dimensions at late times. Figures~\ref{fig:st}(a) and~\ref{fig:st}(b) show the automaton time evolution of the spin autocorrelation function for $p=0$ and $p=0.1$, respectively. The spin dynamics with $p=0$ displays a large anomalous regime where $S(t)$ deviates significantly from $S(t)\sim t^{-1}$; the spin diffusion is very slow at relatively short times ($t<10^{2}$), then becomes faster at intermediate times ($t>10^{2}$), and then asymptotically approaches the diffusive scaling $t^{-1}$. This behavior, including the crossover time scale ($t\sim10^{2}$), is size independent; see Supplemental material~\cite{supp}. By contrast, the spin dynamics with $p=0.1$ approaches the diffusive scaling at much earlier times.

To study momentum-resolved contributions to the spin autocorrelation function, we evaluate the structure factor in momentum and time domain
\begin{align}
    S_{\ell,\ell'}(\bm{k},t) &= \frac{1}{N/3}\sum_{\bm{r}\in I_{\ell}}\sum_{\bm{r}'\in I_{\ell'}}S_{\bm{r},\bm{r}'}(t)\mathrm{e}^{-i\bm{k}\cdot(\bm{r}-\bm{r}')}.
\end{align}
The spin autocorrelation function can be written as the $\bm{k}$-summation of the time-dependent Fourier component $S(t)=(1/N)\sum_{\bm{k}}\mathrm{tr}[S(\bm{k},t)]$. We focus on the momentum-resolved contributions from the vicinity of the high-symmetry points, $S_{\Gamma}(t)=(1/N)\sum_{\bm{k}\sim\bm{0}}\mathrm{tr}[S(\bm{k},t)]$ and $S_{K\cup K'}(t)=(1/N)\sum_{\bm{k}\sim\bm{K},\bm{K}'}\mathrm{tr}[S(\bm{k},t)]$, which dominate at late times as shown in Fig.~\ref{fig:st}. For conventional diffusion dynamics, the dominant contribution comes from near $\Gamma$ point with the form $\mathrm{e}^{-D\bm{k}^{2}t}$. In our system with $p=0$, however, there are also subdominant contributions from the vicinity of $K$ and $K'$ points, which shift the intermediate-time diffusive transport off the diffusive scaling. These contributions decay as $t$ increases and the spin autocorrelation function asymptotically reaches $t^{-1}$, originating from the contributions near $\Gamma$ point. For $p=0.1$, the violation of the spin conservation law on the dynamic sublattices yields an exponential decay of the subdominant contributions at $\bm{k}\sim\bm{K},\bm{K}'$.

The additional contribution $S_{K\cup K'}(t)$ and the broad spectral distribution near the $\tilde{K}'$ point in $S(\bm{k},\omega)$ arises from the spin conservation law on the dynamic sublattices. Although the position of the dynamic sublattice changes during the time evolution, it on average prefers the particular spatial profile determined by the wave vectors $\bm{K}$ and $\bm{K}'$ (see Supplemental material). This creates an effective spatially modulated symmetry of spins and leads to finite $S_{K\cup K'}(t)$. Spatially modulated symmetries refer to conserved quantities that are modulated in space, that give rise to slow dynamics at finite momenta~\cite{sala2022prl, Hart2022}. While the autocorrelation function discussed here contains a sum over the sublattices, the dynamical spin structure factor measures the scattering cross section. Thus the three kagome sublattices lead to distinct contributions in the extended Brillouin zone shown in Fig.~\ref{fig:sdsf} (b).  The slow modes of the autocorrelation near the $K$, $K'$ points then arise in scattering experiments most pronouncedly near the $\tilde K$, $\tilde K'$ points, as shown in Fig.~\ref{fig:sdsf} (c)--(e).

\emph{\textbf{Conclusions and outlook.---}}In the strong-coupling limit, the extended Hubbard model on the kagome lattice with density $2/3$  exhibits exotic dynamical constraints. We have shown that at elevated temperatures these constraints lead to unconventional charge and spin relaxation dynamics, manifesting most prominently as a strong spectral response of the spin correlations at the $\tilde{K}'$ point. The emergent constraints give rise to specific spectral fingerprints, which can be experimentally detected with inelastic neutron scattering experiments.

There are several candidate materials for exploring these unconventional relaxation dynamics. In a strongly correlated organic material Cu-dicyanoanthracene (Cu-DCA), DCA molecules form the kagome lattice with desired density of $n=2/3$~\cite{pawin2008acie,zhang2014cc,zhang2016nl,fuchs2020jpm}. Although $J$ cannot be zero because $t>0$, this material is still expected to show the unconventional charge relaxation.  In a hole-doped herbertsmithite ACu$_{3}$(OH)$_{6}$Cl$_{2}$ ($A=$Li$^{+}$,Na$^{+}$), Cu ion forms the kagome lattice layer with $n=2/3$ and  first-principle calculations predict $t<0$~\cite{guterding2016scirep}. Applying the chemical or hydrostatic pressure to these materials may lead to the desirable parameter range to achieve $J\sim 0$. A small next-nearest-neighbor hopping present in these material classes does not smear out the unconventional spin dynamics, see Supplemental material~\cite{supp}. Mo$_{3}$O$_{8}$ systems such as Li$_{2}X$Mo$_{3}$O$_{8}$ ($X=$Sc, In)~\cite{torardi1985ic} can be described by the extended Hubbard model on the kagome lattice with $n=1/3$~\cite{nikolaev2021npj}. Since the electrons in the strong-coupling regime show a similar constraint dynamics, we expect unconventional relaxation dynamics to prevail in these materials. Various other frustrated lattice models may give rise to related features in inelastic scattering experiments, including for example spin ice systems, strongly correlated electrons on pyrochlore lattices, and other models that effectively map on loop or dimer manifolds~\cite{moessner2000prl,moessner2001prb,moessner2003prb2}.

Our cellular automaton circuit approach is expected to have a wide range of applications in analyzing transport phenomena in materials at elevated temperatures. Moreover, it would be interesting to extend our study of cellular automaton circuit dynamics to models with fracton constraints that are relevant for certain materials~\cite{Yan2020,Han2022} to test for the signatures of their hydrodynamic response~\cite{feldmeier2020prl, gromov2020prr}. Another intriguing future research direction is to investigate monopole dynamics in spin-ice materials~\cite{hallen2022science}.

\emph{\textbf{Acknowledgements.---}}We acknowledge support from the  Deutsche Forschungsgemeinschaft (DFG, German Research Foundation) under Germany's Excellence Strategy--EXC--2111--390814868, DFG grants No. KN1254/1-2, KN1254/2-1, and TRR 360 - 492547816 and from the European Research Council (ERC) under the European Unions Horizon 2020 research and innovation programme (Grant Agreements No. 771537 and No. 851161), as well as the Munich Quantum Valley, which is supported by the Bavarian state government with funds from the Hightech Agenda Bayern Plus.
M.Ka. was supported by JSPS Overseas Research Fellowship.

\emph{\textbf{Data and materials availability.---}}Data analysis and simulation codes are available on Zenodo upon reasonable request~\cite{zenodo}.

\bibliography{biblio}

\newpage
\leavevmode \newpage
\onecolumngrid
\appendix
\begin{center}
\textbf{Supplemental Material\\ \papertitle}\\
\vspace{10pt}
Masataka Kawano$^{1,2}$, and Frank Pollmann$^{1,2}$, Michael Knap$^{1,2}$\\
\vspace{5pt}
$^1$\textit{\small{\tum}}\\
$^2$\textit{\small{\mcqst}}\\
\vspace{10pt}
\end{center}
\twocolumngrid

\section{Cellular automaton circuit}
\begin{figure}[t]
    \includegraphics[width=85mm]{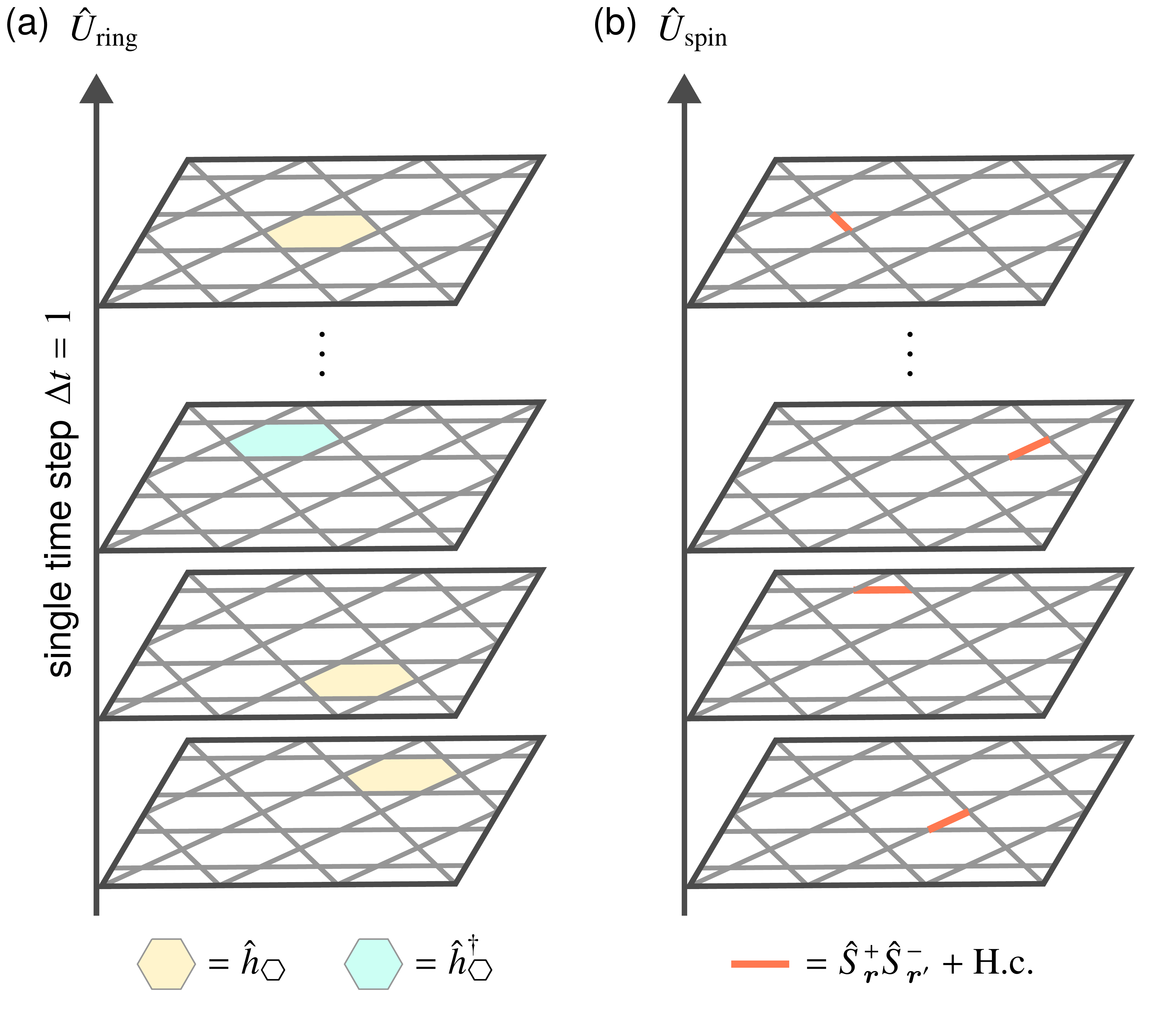}
    \caption{\textbf{Cellular automaton circuit.} Single time step of the cellular automaton circuit consisting of (a) local ring-exchange processes and (b) spin flips. All local updates appear exactly once in a single time step, and the sequence of applying these updates changes in each individual time step.}
    \label{fig:circuit}
\end{figure}

We study the dynamics of charge and spin excitations using a cellular automaton circuit, which encodes both ring-exchange $\hat{U}_{\mathrm{ring}}$ and spin-exchange $\hat{U}_{\mathrm{spin}}$ terms; see Fig.~\ref{fig:circuit}(a) and~\ref{fig:circuit}(b). The unitary operator $\hat{U}_{\mathrm{ring}}$ consists of $N_{\mathrm{h}}$, the number of hexagons in the kagome lattice, local ring-exchange updates in a clockwise ($\hat{h}_{\hexagon}$) or counterclockwise ($\hat{h}_{\hexagon}^{\dagger}$) orientation. For a local update, either $\hat{h}_{\hexagon}$ or $\hat{h}_{\hexagon}^{\dagger}$ is chosen with a probability $1/2$, and the order of local updates is altered for each individual time step. The unitary operator $\hat{U}_{\mathrm{spin}}$ is constructed similarly; all local spin flips $\hat{S}_{\bm{r}}^{+}\hat{S}_{\bm{r}'}^{-}+\mathrm{H.c.}$ appear exactly once during a single time step in random order. These operators encode the relevant structure of the strong coupling Hamiltonian~\eqref{eq:Hring} and \eqref{eq:Hspin}. One can also consider the stochastic automaton, in which the local updates are applied with a finite acceptance probability. We have checked that the implementation details, including the acceptance probability, merely change the timescale and do not influence the structure of the dynamics.

The cellular automaton time evolution $\hat{U}(t)$ is driven by the sequential application of $\hat{U}_{\mathrm{ring}}$ and $\hat{U}_{\mathrm{spin}}$, e.g. after six steps the evolution could be $\hat{U}_{\mathrm{ring}}\hat{U}_{\mathrm{ring}}\hat{U}_{\mathrm{ring}}\hat{U}_{\mathrm{spin}}\hat{U}_{\mathrm{ring}}\hat{U}_{\mathrm{ring}}$. The effect of finite $J/g$ is reflected as the finite probability $p$ of applying $\hat{U}_{\mathrm{spin}}$ at each time step. Due to this construction, the application of $\hat{U}(t)$ to an initial product state $\ket{\psi(0)}$ leads to another produce state $\ket{\psi(t)}=\hat{U}(t)\ket{\psi(0)}$ at time $t$. In the regime $U,V\gg k_{\mathrm{B}}T\gg t$, the Boltzmann weight is almost identical to $1/\mathrm{dim}\mathbb{H}$ in the low-energy sector, where $\mathbb{H}$ is the low-energy Hilbert space we consider. While the Boltzmann weight in the high-energy sector is negligibly small. The spin correlation function $S_{\bm{r},\bm{r}'}(t)$ is then calculated as
\begin{align}
    S_{\bm{r},\bm{r}'}(t) &= \frac{1}{\dim\mathbb{H}}\sum_{\ket{\psi}\in\mathbb{H}}\braket{\psi|\hat{U}^{\dagger}(t)\hat{S}_{\bm{r}}^{z}\hat{U}(t)\hat{S}_{\bm{r}'}|\psi} \nonumber \\
                          &= \frac{1}{\dim\mathbb{H}}\sum_{\ket{\psi}\in\mathbb{H}}S_{\bm{r}}^{z}(\psi(t))S_{\bm{r}}^{z}(\psi(0)),
    \label{eq:scorr_full}
\end{align}
where $S_{\bm{r}}^{z}(\psi(t))$ satisfies $\hat{S}_{\bm{r}}^{z}\ket{\psi(t)}=S_{\bm{r}}^{z}(\psi(t))\ket{\psi(t)}$. Since the full summation of $\ket{\psi}$ for large systems is immensely difficult, we replace the summation in Eq~(\ref{eq:scorr_full}) with the sample average
\begin{align}
    S_{\bm{r},\bm{r}'}(t) \simeq \frac{1}{M}\sum_{\ket{\psi}}S_{\bm{r}}^{z}(\psi(t))S_{\bm{r}}^{z}(\psi(0)),
\end{align}
where $M$ is the number of samples. The charge correlation function can also be calculated in a similar manner. Since every charge configuration can be mapped to the fully packed loops on the dual honeycomb lattice~\cite{pollmann2014prb}, we can use the worm algorithm~\cite{rahman1972jcp,evertz1993prl}, which is known to be ergodic, to sample the initial charge configurations. The sampling of spins involves randomly assigning $1/2$ or $-1/2$ to the charge with probability $1/2$.
To evaluate the dynamic structure factors, we replace the integral $\int_{0}^{\infty}dt$ with the discrete summation $\sum_{t=0}^{t_{\mathrm{max}}}$ as
\begin{align}
    C(\bm{k},\omega)&\simeq\frac{2}{N}\sum_{t=0}^{t_{\mathrm{max}}}\mathrm{Re}\left[C_{\bm{r},\bm{r}'}(t)\mathrm{e}^{-i\bm{k}\cdot(\bm{r}-\bm{r}')+i\omega t}\right],\\
    S(\bm{k},\omega)&\simeq\frac{2}{N}\sum_{t=0}^{t_{\mathrm{max}}}\mathrm{Re}\left[S_{\bm{r},\bm{r}'}(t)\mathrm{e}^{-i\bm{k}\cdot(\bm{r}-\bm{r}')+i\omega t}\right].
\end{align}

We can also construct the cellular automaton circuit from Trotter gates, which clarifies the relationship between $p$ and $J/g$. We first focus on the local ring-exchange process in $\hat{U}_{\mathrm{ring}}$, namely $\ket{\psi'}=(\hat{h}_{\hexagon}+\mathrm{H.c.})\ket{\psi}$. This process can be written as the product of $N_{g}=\pi/(2|g|\Delta t)$ Trotter gates as $\ket{\psi'}\propto[\hat{U}_{g}(\Delta t)]^{N_{g}}\ket{\psi}$ with $\hat{U}_{g}(\Delta t)=\exp[-ig(\hat{h}_{\hexagon}+\mathrm{H.c.})\Delta t]$ Similarly, the local spin-flip process in $\hat{U}_{\mathrm{spin}}$ can be written as the product of $N_{J}=\pi/(|J|\Delta t)$ Trotter gates $\hat{U}_{J}(\Delta t)=\mathrm{exp}[-i(J/2)(\hat{S}_{\bm{r}}^{+}\hat{S}_{\bm{r}'}^{-}+\mathrm{H.c.})\Delta t]$. Since $\exp[-i\hat{\mathcal{H}}_{\mathrm{eff}}\Delta t]\simeq\exp[-i\hat{\mathcal{H}}_{\mathrm{ring}}\Delta t]\exp[-i\hat{\mathcal{H}}_{\mathrm{spin}}\Delta t]$ for $\Delta t\ll1$, the two Trotter gates, $\hat{U}_{g}(\Delta t)$ and $\hat{U}_{J}(\Delta t)$, appear in equal numbers during the time evolution. By rearranging the Trotter gates, we obtain the cellular automaton circuit consisting of $n_{g}=N_{t}/N_{g}$ ring exchange $\hat{U}_{\mathrm{ring}}$ and $n_{J}=N_{t}/N_{J}$ spin exchange $\hat{U}_{\mathrm{spin}}$, where $N_{t}=t/\Delta t$. Therefore, the probability of having $\hat{U}_{\mathrm{spin}}$ at each time step can be estimated by $p \sim n_{J}/(n_{g}+n_{J})=|J|/(2|g|+|J|)$, which is determined by the ratio $|J/g|$. For $|J/g|\ll1$, the dynamics of the system is effectively described by the cellular automaton circuit with small $p\sim|J/(2g)|\ll1$.

\section{Effective field theory for charge dynamics}
In the main text, we have discussed how any charge configuration of the system can be mapped onto a loop model on a dual honeycomb lattice. Then we can assign a height variable $z_{\bm r}$ to each of the hexagons. The charge density $\delta n_{\bm{r}}$ can be written in terms of the height variable as~\cite{moessner2003prb} $\delta n_{\bm{r}}=(1/3)(z_{\bm{r}+\bm{e}_{\ell}/2}-z_{\bm{r}-\bm{e}_{\ell}/2})$ for $\bm{r}\in I_{\ell}$, where $\bm{e}_{1}=(1,0)$, $\bm{e}_{2}=(-1/2,\sqrt{3}/2)$, and $\bm{e}_{3}=(-1/2,-\sqrt{3}/2)$ are vectors connecting next-nearest neighbor sites on the dual honeycomb lattice. We also define an average value of the height defined on the dual honeycomb lattice sites as $h_{\bm{r}}=(1/3)(z_{\bm{r}_{1}}+z_{\bm{r}_{2}}+z_{\bm{r}_{3}})$, where $\bm{r}_{i}$ ($i=1,2,3$) are the positions of neighboring hexagons around $\bm{r}$.

As a next step, we need to coarse grain the height field and take the continuum limit: $h_{\bm{r}}\to h(\bm{r})$. The key observation is that the height variable $z_{\bm{r}}$, and therefore the charge density $\delta n_{\bm{r}}$, are rapidly modulated in space with the wave vector $\bm{K}$ and $\bm{K}'$ even when $h(\bm{r})$ takes a uniform value~\cite{zeng1997prb}. The continuum limit of the charge density, $\delta n_{\bm{r}}\to\delta n_{\ell}(\bm{r})$, should respect this spatial structure, and can be written in terms of the average height $h(\bm{r})$ as~\cite{moessner2003prb,fradkin2004prb}
\begin{align}
    \delta n_{\ell}(\bm{r}) = \frac{1}{3}\partial_{\ell}h(\bm{r}) &+\frac{1}{2}\cos\left(\bm{K}\cdot\bm{r}+\phi_{\ell}+\frac{2\pi}{3}h(\bm{r})\right) \nonumber \\
                                                                  &+\frac{1}{2}\cos\left(\bm{K}'\cdot\bm{r}-\phi_{\ell}-\frac{2\pi}{3}h(\bm{r})\right),
    \label{eq:continuum_n}
\end{align}
where $\partial_{\ell}=\bm{\nabla}\cdot\bm{e}_{\ell}$ comes from the relation $\delta n_{\bm{r}}=(1/3)(z_{\bm{r}+\bm{e}_{\ell}/2}-z_{\bm{r}-\bm{e}_{\ell}/2})$ and $\phi_{\ell}=2(\ell-1)\pi/3$ reflects the three sublattice structure of the kagome lattice. The height variable in the second and third terms appears as $(2\pi/3)h(\bm{r})$, reflecting the local loop configurations. Therefore the particle density is invariant under a shift of the height by $3$~\cite{henley1997jsp}.

We assume that the dynamics of the height is governed by the action~\cite{henley1997jsp}
\begin{align}
    S[h(\bm{r})] = \frac{\rho}{2}\int d^{2}\bm{r}\left(\bm{\nabla}h(\bm{r})\right)^{2},
\end{align}
with $\rho=\pi/9$, which is chosen to reproduce the asymptotic behavior of the correlation function of the lattice system. This action reflects the fact that the configurations with uniformly distributed $h(\bm{r})$ are typical states and the system relaxes to these states due to the entropic effect. We now assume that the dynamics of the height is governed by the Langevin equation~\cite{henley1997jsp}
\begin{align}
    \frac{\partial}{\partial t} h(\bm{r},t) = -\gamma\frac{\delta S[h(\bm{r},t)]}{\delta h(\bm{r},t)} + \zeta(\bm{r},t),
    \label{eq:langevin}
\end{align}
where $\gamma$ is proportional to the diffusion constant and has to be fitted to the cellular automaton dynamics once, as it sets the timescale of the dynamics. The first and second terms of the right-hand side in Eq.~(\ref{eq:langevin}) represents the damping force and randomly fluctuating force satisfying $\braket{\zeta(\bm{r},t)\zeta(\bm{r}',t')}=2\gamma\delta(\bm{r}-\bm{r}')\delta(t-t')$. The time-dependent height correlation function $\braket{h(\bm{r},t)h(\bm{r}',0)}$ then follows a diffusion equation with diffusion constant $\rho\gamma$, and we have
\begin{align}
    \braket{h(\bm{r},t)h(\bm{0},0)} = \braket{\bar{h}(t)\bar{h}(0)} + \frac{1}{V}\sum_{\bm{q}\neq\bm{0}}\frac{\mathrm{e}^{-\rho\gamma\bm{q}^{2}t}}{\rho\bm{q}^{2}}\cos\bm{q}\cdot\bm{r},
    \label{eq:hcorr}
\end{align}
where $V=\sqrt{3}N_{\mathrm{h}}/2$ is the volume, $\bar{h}(t)=(1/V)\int d^{2}\bm{r}\ h(\bm{r},t)$ is the average height, and discrete summation $\sum_{\bm{q}\neq\bm{0}}$ runs over the first Brillouin zone of the kagome lattice with $N_{\mathrm{h}}$ hexagons, naturally acting as the momentum cutoff~\cite{henley1997jsp}.

The charge correlation function is calculated from Eqs.~(\ref{eq:continuum_n}) and~(\ref{eq:hcorr}) as
\begin{align}
    \frac{1}{3}\sum_{\ell}\braket{\delta n_{\ell}(\bm{r},t)\delta n_{\ell}(\bm{0},0)} &= \frac{1}{9\rho V} + \frac{1}{18\rho V}\sum_{\bm{q}\neq\bm{0}}\mathrm{e}^{-\rho\bm{q}^{2}\gamma t}\cos\bm{q}\cdot\bm{r} \nonumber \\
                                                                                      &+\frac{1}{8}\sum_{\bm{k}_{0}=\pm\bm{K},\pm\bm{K}'}\mathrm{e}^{i\bm{k}_{0}\cdot\bm{r}-(4\pi^{2}/9)[g(\bm{r},t)-\bar{g}(t)]}.
    \label{eq:ccorr}
\end{align}
where we introduce $g(\bm{r},t)=\braket{[h(\bm{r},t)-h(\bm{0},0)]^{2}/2}$ and $\bar{g}(t)=\braket{[\bar{h}(t)-\bar{h}(0)]^{2}/2}$, which are calculated as
\begin{align}
    g(\bm{r},t) &= \frac{1}{V}\sum_{\bm{q}\neq\bm{0}}\frac{1}{\rho\bm{q}^{2}}\left(1-\mathrm{e}^{-\rho\bm{q}^{2}\gamma t}\cos\bm{q}\cdot\bm{r}\right),
    \label{eq:g(r,t)} \\
    \bar{g}(t) &= \frac{\gamma t}{V}.
    \label{eq:g(t)}
\end{align}
The dynamics of $\bar{h}(t)$ leads to the exponential decay of the second term in Eq.~(\ref{eq:ccorr}), arising from the diffusion of the height variable in the height space of periodicity $3$ with diffusive constant $\gamma/2V$~\cite{henley1997jsp}. We also add the constant term $1/9\rho V$ in Eq.~(\ref{eq:ccorr}), which comes from the charge conservation law. The charge autocorrelation function and time-dependent Fourier component can be calculated from Eq.~(\ref{eq:ccorr}) as
\begin{align}
    C(t) = \frac{1}{3}\sum_{\ell}\braket{\delta n_{\ell}(\bm{0},t)\delta n_{\ell}(\bm{0},0)},
    \label{eq:autocorr}
\end{align}
\begin{align}
    \mathrm{tr}[C(\bm{k},t)] = \sum_{\bm{r}}\sum_{\ell}\braket{\delta n_{\ell}(\bm{r},t)\delta n_{\ell}(\bm{0},0)}\mathrm{e}^{-i\bm{k}\cdot\bm{r}}.
    \label{eq:tdcsf}
\end{align}
Here, we use a discrete summation with respect to $\bm{r}$ in Eq.~(\ref{eq:tdcsf}) by considering the triangular lattice formed by hexagons where the height variable $z_{\bm{r}}$ lives. This hight-field theory reproduces the automaton dynamics extremely well, except at a single point, the $K$ point, in momentum space, see main text for a discussion.

\section{Role of dynamic sublattice constraint}
\begin{figure}[t]
    \includegraphics[width=85mm]{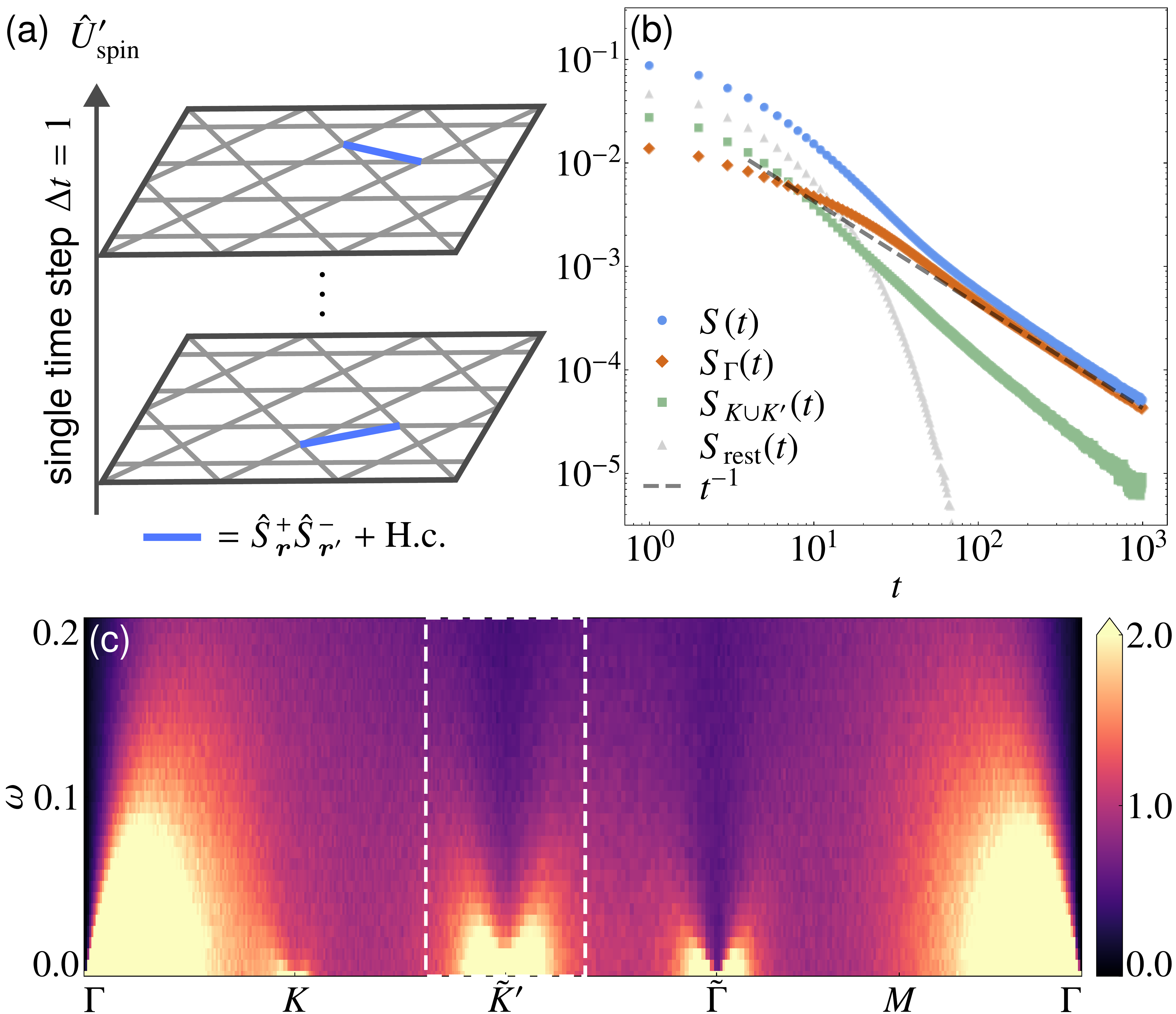}
    \caption{\textbf{Spin relaxation dynamics with next-nearest-neighbor spin exchange interactions for $\bm{p'=0.1}$.} (a) Single time step of the cellular automaton circuit consisting of next-nearest-neighbor spin exchanges. (b) Time evolution of the spin autocorrelation function $S(t)$, together with the momentum-resolved contributions, $S_{\Gamma}(t)$, $S_{K\cup K'}(t)$, and $S_{\mathrm{rest}}(t)$, defined in the main text. (c) Dynamic spin structure factor $S(\bm{k},\omega)$ along the cut through the Brillouin zone shown in Fig.~\ref{fig:sdsf}(b). We still observe a strong spectral response near the $\tilde{K}'$ point (enclosed by the white dots). We use $N_{\mathrm{h}}=150\times150$ systems with the periodic boundary condition and $2\times10^{5}$ samples.}
    \label{fig:tracer}
\end{figure}

To further support that the spin conservation law on the dynamic sublattices is responsible for the strong spectral features at the $\tilde K'$ point of the dynamical spin structure factor, we introduce an additional next-nearest-neighbor spin exchange term $\hat{U}_{\mathrm{spin}}'$ shown in Fig.~\ref{fig:tracer}(a).  This term preserves the spin conservation law on the dynamic sublattices but induces further nontrival spin dynamics. We denote the probability of having $\hat{U}_{\mathrm{spin}}'$ as $p'$ and consider $p'=0.1$ here. The time evolution of the spin autocorrelation function and momentum-resolved contributions obtained by the cellular automaton circuit simulations consisting of $\hat{U}_{\mathrm{ring}}$ and  $\hat{U}_{\mathrm{spin}}'$ show again anomalous diffusive scaling due to strong contributions from $S_{K\cup K'}(t)$; see Fig.~\ref{fig:tracer}(b). The dynamic spin structure factor again exhibits high spectral intensity near the $\tilde{K}'$ point; see Fig.~\ref{fig:tracer}(c). This further confirms that the unconventional spin transport in our system arises from the spin conservation laws on the dynamic sublattices.

\section{Effect of the next-nearest-neighbor hopping}
To examine the robustness of the unconventional spin relaxation dynamics, we consider an additional next-nearest-neighbor hopping $\mathfrak{t}'$ in the Hamiltonian (\ref{eq:H}). The lowest order effective Hamiltonian is given by $\hat{\mathcal{H}}_{\mathrm{eff}} = \hat{\mathcal{H}}_{\mathrm{ring}} + \hat{\mathcal{H}}_{\mathrm{spin}} + \hat{\mathcal{H}}_{\mathrm{spin}}'$. The coupling constant $g$ and $J$ in $\hat{\mathcal{H}}_{\mathrm{ring}}$ and $\hat{\mathcal{H}}_{\mathrm{spin}}$ are the same as the ones defined in the main text. The additional term $\hat{\mathcal{H}}_{\mathrm{spin}}'$ consists of the next-nearest-neighbor and next-nearest-neighbor exchange interactions,
\begin{align}
    \hat{\mathcal{H}}_{\mathrm{spin}}' &= J' \sum_{\braket{\braket{\bm{r},\bm{r}'}}} \left( \hat{\bm{S}}_{\bm{r}}\cdot\hat{\bm{S}}_{\bm{r}'} - \frac{1}{4}\hat{n}_{\bm{r}}\hat{n}_{\bm{r}'} \right) \nonumber \\
                                       &\hspace{10pt} + \tilde{J}\sum_{\braket{\bm{r},\bm{r}'}} \left( \hat{\bm{S}}_{\bm{r}}\cdot\hat{\bm{S}}_{\bm{r}'} - \frac{1}{4}\hat{n}_{\bm{r}}\hat{n}_{\bm{r}'} \right)\sum_{\underset{\bm{r},\bm{r}',\bm{r}''\in\triangle_{\mathfrak{t}\mathfrak{t}\mathfrak{t}'}}{\bm{r}''}}(1-\hat{n}_{\bm{r}''}) \nonumber \\
                                       &\hspace{10pt} + \tilde{J}_{1}'\sum_{\braket{\braket{\bm{r},\bm{r}'}}} \left( \hat{\bm{S}}_{\bm{r}}\cdot\hat{\bm{S}}_{\bm{r}'} - \frac{1}{4}\hat{n}_{\bm{r}}\hat{n}_{\bm{r}'} \right)\sum_{\underset{\bm{r},\bm{r}',\bm{r}''\in\triangle_{\mathfrak{t}\mathfrak{t}\mathfrak{t}'}}{\bm{r}''}}(1-\hat{n}_{\bm{r}''}) \nonumber \\
                                       &\hspace{10pt} + \tilde{J}_{2}'\sum_{\braket{\braket{\bm{r},\bm{r}'}}} \left( \hat{\bm{S}}_{\bm{r}}\cdot\hat{\bm{S}}_{\bm{r}'} - \frac{1}{4}\hat{n}_{\bm{r}}\hat{n}_{\bm{r}'} \right)\sum_{\underset{\bm{r},\bm{r}',\bm{r}''\in\triangle_{\mathfrak{t}'\mathfrak{t}'\mathfrak{t}'}}{\bm{r}''}}(1-\hat{n}_{\bm{r}''})
   \label{eq:H_spin'}
\end{align}
where $\braket{\braket{\bm{r},\bm{r}'}}$ denotes the pairs of next-nearest-neighboring sites, $\triangle_{\mathfrak{t}\mathfrak{t}\mathfrak{t}'}$ denotes the triangle consisting of two nearest-neighbor bonds and one next-nearest-neighbor bond, and $\triangle_{\mathfrak{t}'\mathfrak{t}'\mathfrak{t}'}$ denote the triangle consisting of three next-nearest-neighbor bonds. The coupling constants in Eq.~(\ref{eq:H_spin'}) are given by $J'=4\mathfrak{t}'^{2}/U$, $\tilde{J}=6\mathfrak{t}^{2}\mathfrak{t}'/[V(U-V)]+2\mathfrak{t}^{2}\mathfrak{t}'/V^{2}$, $\tilde{J}_{1}'=8\mathfrak{t}^{2}\mathfrak{t}/UV+4\mathfrak{t}^{2}\mathfrak{t}'/V^{2}$, and $\tilde{J}_{2}'=4\mathfrak{t}'^{3}/UV+\mathfrak{t}'^{3}/V^{2}$. Among these terms, only the nearest-neighbor exchange $\tilde{J}$ violates the spin conservation law on the dynamic sublattice. Because of the additional factor $(1-\hat{n}_{\bm{r}''})$, we cannot eliminate the nearest-neighbor exchange interaction by setting $J+\tilde{J}=0$. However, when $J=0$, the ratio $|\tilde{J}/g|$ is written as $|\tilde{J}/g|=|\mathfrak{t}'/3\mathfrak{t}|(1+2V/U)/(1-V/U)$, which is roughly proportional to $|\mathfrak{t}'/\mathfrak{t}|$ and is expected to be small in candidate materials Cu-DCA~\cite{zhang2016nl,fuchs2020jpm} and ACu$_{3}$(OH)$_{6}$Cl$_{2}$ ($A=$Li$^{+}$,Na$^{+}$)~\cite{guterding2016scirep}. Since the small nearest-neighbor spin exchange does not affect the spectral response much as shown in Fig.~\ref{fig:sdsf}(c-e), we still expect the unconventional spin relaxation dynamics in the materials with small $\mathfrak{t}'$.

\section{Average position of the dynamic sublattice}
\begin{figure}[t]
    \includegraphics[width=85mm]{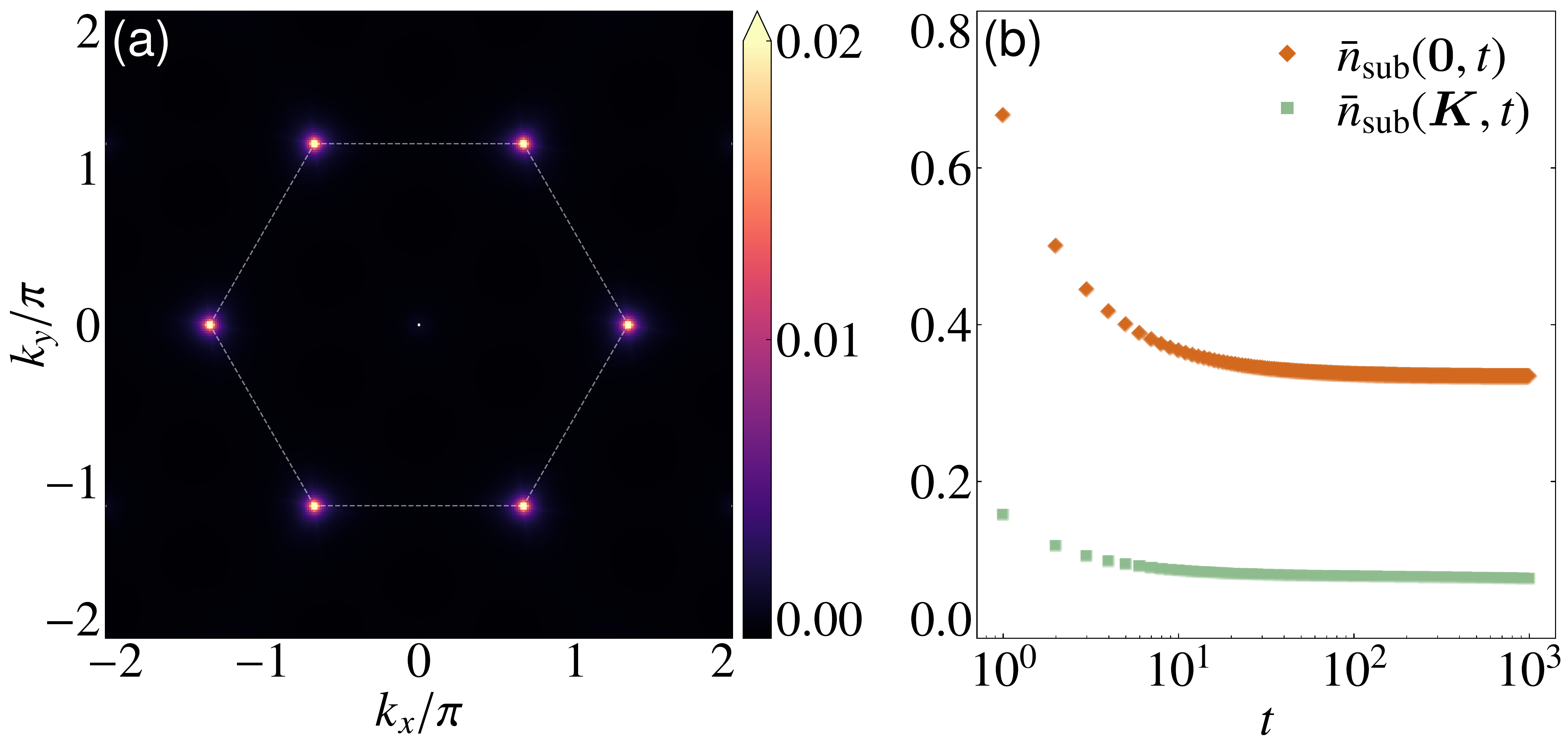}
    \caption{\textbf{Average position of the dynamic sublattice.} (a) Profile of $\bar{n}_{\mathrm{sub}}(\bm{k},t)$ in momentum space at $t=100$ for $N_{h}=150\times150$. (b) Time evolution of $\bar{n}_{\mathrm{sub}}(\bm{k},t)$ at $\bm{k}=0$ and $\bm{K}$.}
    \label{fig:sub}
\end{figure}

To confirm that the position of the dynamic sublattice prefers the spatial profile characterized by the wave vectors $\bm{K}$ and $\bm{K}'$ on average, we focus on the time-averaged charge density on the dynamic sublattice $X$ ($=$A,B)
\begin{align}
    \bar{n}_{\bm{r},X}^{\psi}(t)=\frac{1}{t}\sum_{s=0}^{t}n_{\bm{r},X}(\psi(s)),
\end{align}
where $n_{\bm{r},X}(\psi(s))$ is the charge density on the dynamic sublattice $X$ for product state $\ket{\psi(s)}$. We also introduce the Fourier transformation of $\bar{n}_{\bm{r},X}^{\psi}(t)$ as
\begin{align}
    \bar{n}_{X,\ell}^{\psi}(\bm{k},t) = \frac{1}{N/3}\sum_{\bm{r}\in I_{\ell}}\bar{n}_{\bm{r},X}^{\psi}(t)\mathrm{e}^{-i\bm{k}\cdot\bm{r}}.
\end{align}
The peak position of $\bar{n}_{X,\ell}^{\psi}(\bm{k},t)$ in momentum space characterizes the spatial profile of the time-averaged position of the dynamic sublattice. For concreteness, we consider the sublattice average $\bar{n}_{\mathrm{sub}}^{\psi}(\bm{k},t)=(1/6)\sum_{X=\mathrm{A,B}}\sum_{\ell=1}^{3}|\bar{n}_{X,\ell}^{\psi}(\bm{k},t)|$ and its sample average
\begin{align}
    \bar{n}_{\mathrm{sub}}(\bm{k},t) = \frac{1}{\mathrm{dim}\mathbb{H}}\sum_{\ket{\psi}\in\mathbb{H}}\bar{n}_{\mathrm{sub}}^{\psi}(\bm{k},t).
\end{align}
Figure~\ref{fig:sub}(a) shows the momentum-space profile of $\bar{n}_{\mathrm{sub}}(\bm{k},t)$ at $t=100$. In addition to a dominant peak at $\Gamma$ point, there are two peaks at $K$ and $K'$ points. Therefore, the dynamic sublattice shows the spatially modulated profile determined by $\bm{K}$ and $\bm{K}'$. The time evolution of $\bar{n}_{\mathrm{sub}}(\bm{k},t)$ at $\Gamma$ and $K$ points are shown in Fig.~\ref{fig:sub}(b). Both $\bar{n}_{\mathrm{sub}}(\bm{0},t)$ and $\bar{n}_{\mathrm{sub}}(\bm{K},t)$ remains finite even at late times.

\newpage
\section{System-size dependence of the autocorrelation functions}
\begin{figure}[h]
    \includegraphics[width=85mm]{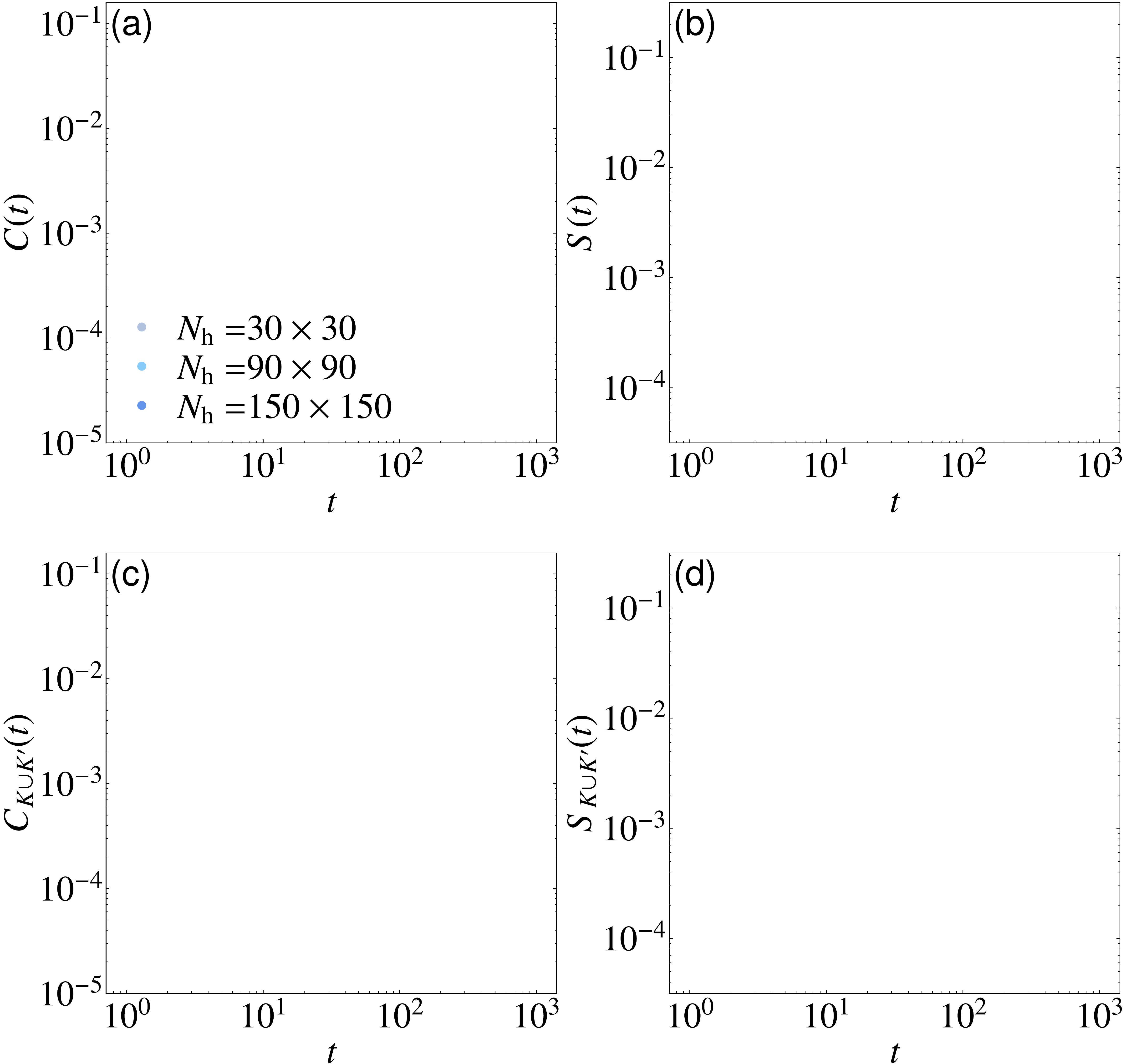}
    \caption{\textbf{System-size dependence of the autocorrelation functions.} Time evolution of the (a) charge autocorrelation function (b) spin autocorrelation function, (c) $C_{K\cup K'}(t)$, and (d) $S_{K\cup K'}(t)$ for different system sizes $N_{\mathrm{h}}=30\times30,90\times90,150\times150$.}
    \label{fig:size}
\end{figure}

On the time-scales shown in the main text, finite system-size effects are negligible, as can be confirmed from the system-size dependence of the autocorrelation functions, shown in Fig.~\ref{fig:size}.

\end{document}